\documentstyle[12pt,epsf]{article}

\renewcommand{\bar}{\overline}

\begin{document}
\begin{flushright}
SLAC--PUB--7096\\
January 1996
\end{flushright}
\bigskip\bigskip

\centerline{{\large Rapidity Gaps in Deep Inelastic Scattering}
    \footnote{\baselineskip=14pt
     Work partially supported by the Department of Energy, contract 
     DE--AC03--76SF00515.}}
\vspace{22pt}
  \centerline{\bf J. D. Bjorken}
\vspace{8pt}
  \centerline{\it Stanford Linear Accelerator Center}
  \centerline{\it Stanford University, Stanford, California 94309}
\centerline{and}
  \centerline{\it Department of Theoretical Physics}
  \centerline{\it Oxford University, Oxford, United Kingdom}
\vspace*{0.3cm}
\centerline{E-mail:bjorken@slac.stanford.edu}
\vspace*{0.9cm}
\begin{center}
Abstract
\end{center}
A simple semiquantitative picture of diffractive electroproduction
is described.  Although the diffractive component of $F_2$ is
approximately independent of $Q^2$ and $W^2$, this mechanism is
``soft,'' i.e. it depends upon large-distance physics and is not
readily describable within perturbative QCD. 
\vspace*{2.9cm}

\begin{center}
Talk presented at the \\
Conference of Nuclear Physics \\
Fundamental Interactions of Elementary Particles \\
October 23--27, 1995 --- ITEP, Moscow
\end{center}

\newpage

\baselineskip=24pt

\section{Introduction}

The subject of this talk has become of considerable current
interest, as evidenced by the other presentations at this meeting
\cite{wagner,mueller,lipatov} on both experimental and theoretical
developments. The main thrust of this one is to argue that, despite
a growing tendency for experimentalists and theorists to say
otherwise, the presence of a diffractive contribution which at large
$Q^2$ and small $x$ ``scales", i.e. comprises a finite fraction (of
order 5 to 10 percent) of all events, does not imply that the
mechanism which creates this class of events is pointlike.
Furthermore, a long standing mechanism \cite{bj:align} for creating
the observed nondiffractive final-state properties (the so-called
``aligned-jet" mechanism) appears quite sufficient to account for
the bulk of the experimental evidence on diffractive final states.

It is especially appropriate to discuss this at this meeting, the
50th anniversary celebration of ITEP, because for me its origins go
back to what I learned in my visits here in the early 1970s.
Foremost was the very early observation by Ioffe \cite{ioffe} that,
at small $x$ and at large $Q^2$, large longitudinal distances $z$
are important in the spacetime structure of the forward
virtual-photon-proton Compton amplitude, the absorptive part of
which determines the deep-inelastic structure functions. In the
target-proton rest frame the estimate is roughly
\begin{equation}
  z = \frac{1}{M_p~x}\approx \frac{2\nu}{Q^2}\ .
\end{equation}   

Later, Gribov \cite{gribov} used this observation to argue that
generalized vector-meson dominance could provide an estimate of the
behavior of the structure function in the limit of very small $x$
and very large target (e.g. a nucleus). He used a most
straightforward line of reasoning, which however led to a paradox,
indeed a disaster, because the resultant structure function did not
even approximately scale--it was too big at large $Q^2$ by a full
power of $Q^2$. It is this paradox and its resolution which is the
centerpiece of the discussion to follow.

\section{Basic physics of $F_2(x,Q^2)$ at small $x$}

\begin{figure}[htbp]
\begin{center}
\leavevmode
\epsfxsize=5in
\epsfbox{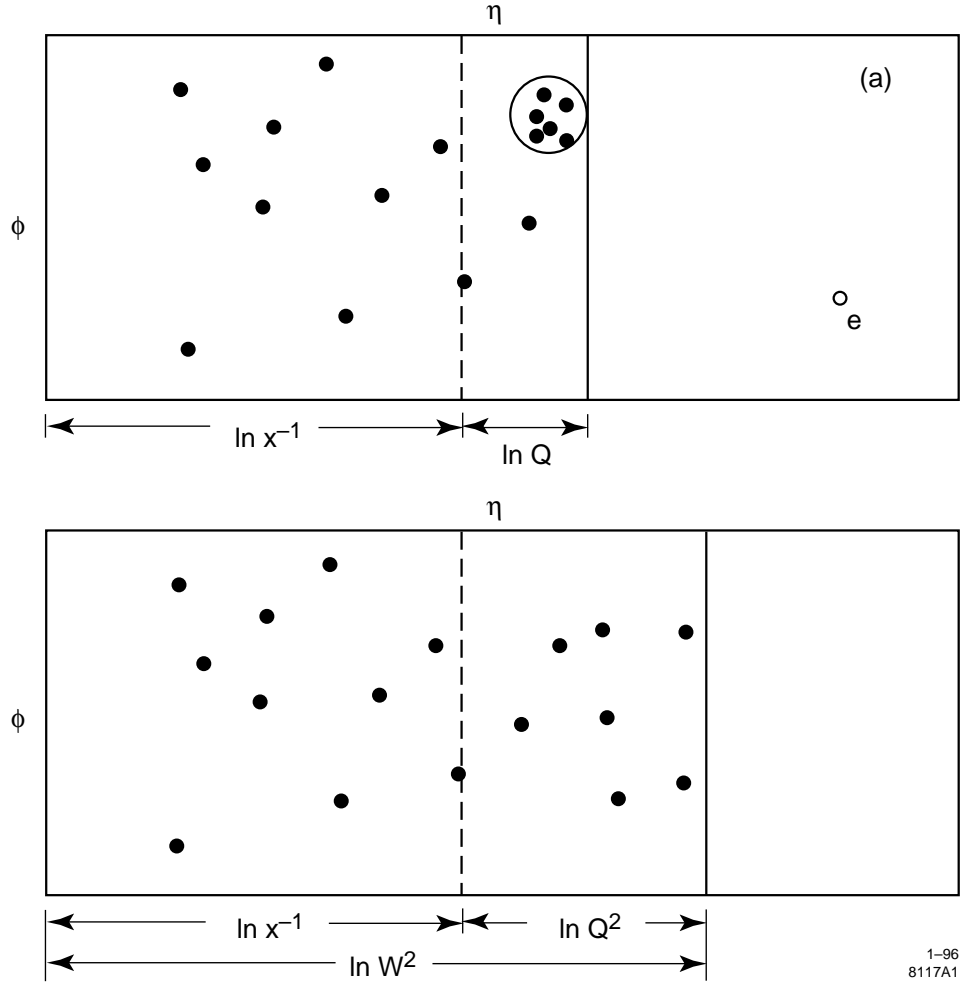}
\end{center}
\caption[*]{Lego-plot of final-state hadrons in small-$x$ deep
inelastic scattering: 
\hfill\break (a) HERA laboratory frame, and (b) collinear
$\gamma^* - p$ reference frames.}
\label{fig1}
\end{figure}

The familiar parton picture of deep-inelastic scattering is easy to
apply at small $x$, especially in a typical HERA laboratory frame of
reference. A right-moving electron ``sees" a left-moving wee parton
in the extreme-relativistic left-moving proton, and Coulomb-scatters
from it with momentum transfer $Q$. The lego-plot picture of the
final-state particles is sketched in Fig. \ref{fig1}a; one sees the
electron and the struck-quark jet each with transverse momentum $Q$.
This Coulomb-scattering picture is accurate in the frame of
reference where $\eta = 0$ is chosen to be the rapidity halfway
between these two features.

Also of interest is the (approximate) location of the initial state
quark before it was struck (the so-called ``hole" fragmentation
region \cite{bj:holes}). It is a distance of order $\ell n\, Q$ to the
left of the quark jet, because $\ell n\, \theta/2$ has changed by
approximately that amount because of the Coulomb scattering. It is a
distance $\ell n\, 1/x$ from the leading-proton fragments.

It is also convenient, especially theoretically, to view the same
process in a collinear virtual-photon proton reference frame (cf.
Fig. \ref{fig1}b). In such a frame there are generically no
large-$p_T$ jets, at least at the level of naive, old-fashioned
parton model. With QCD, there will be extra gluon initial-state and
final-state radiation. Most of this will look like minijet
production in collinear reference frames, but occasionally there
will be extra genuine gluon jets, especially in the phase-space
region between the hole and the leading quark fragments. Note that
now the amount of phase-space to the right of the ``hole" region is
of order $\ell n\, Q^2$; the extra $\ell n\, Q$ amount of phase space is
in the quark jet in HERA reference frames. The total phase space is
evidently $\ell n\, Q^2 + \ell n\, 1/x = \ell n\, W^2$, as it should be.

\vspace{.5cm}

We now are ready to introduce Gribov's paradox. He viewed the same
process in the laboratory frame of the nucleon, but considered for
simplicity replacement of the nucleon by a large, heavy nucleus of
radius $R$. The picture is that first there is the virtual
dissociation of the virtual photon into a hadron system upstream of
the target hadron. For HERA conditions and Ioffe's estimate of
longitudinal distances, this is a distance of hundreds of fermis in
this fixed-target reference frame. This virtual dissociation process
is followed by just the geometrical absorption of the virtual-hadron
system on the nucleus. Gribov used old-fashioned (or in modern terms
light-cone) perturbation theory to make the estimate, which gives a
simple and utterly transparent result:
\begin{equation}
            \sigma_T = (1 - Z_3) \, \pi R^2 \ .
\end{equation}

Note that the estimate is for $\sigma_T$, not $F_2$, and that
$(1 - Z_3)$, with $Z_3$  the charge renormalization of the photon,
is just the probability the photon is hadron, not photon:
\begin{equation}
1 - Z_3 = \frac{\alpha}{3\pi} \int \frac{ds~s~R(s)}{(Q^2+s)^2}\approx
\frac{\alpha}{3\pi}~\bar{R}~\ell n\, \frac{1}{x}\ ,
\end{equation}
where $R(s)$ is the sum of squared charges of partons, as used in
describing the $e^+$-$e^-$ annihilation cross section. So up to
logarithmic factors, the result is that $\sigma_T$, not $F_2$, is
independent of $Q^2$. Since
\begin{equation}
       F_2 = \frac{Q^2~\sigma_T}{4\pi^2~\alpha}\ ,
\end{equation}
this means the aforementioned scaling violation by an extra power of
$Q^2$. Gribov's structure function is much too big (at least at
present energies)!!

\begin{figure}[htbp]
\begin{center}
\leavevmode
\epsfxsize=5in
\epsfbox{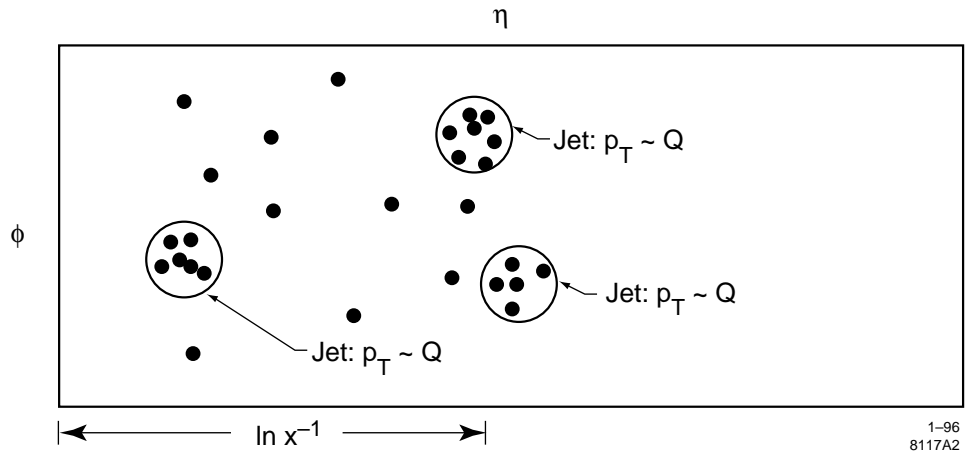}
\end{center}
\caption[*]{Lego-plot of final-state hadrons for the
``color-transparency", ``Bethe-Heitler", or ``noncollinear
photon-gluon fusion" mechanism at small $x$.}
\label{fig2}
\end{figure}

There are (at least) two ways out of the paradox. One way is, in
modern jargon, ``color transparency". Typically the virtual photon
dissociates into a bare $q- \bar{q}$ system which on arrival at the
nucleus is a small color dipole of spatial extent $Q^{-1}$. It can
only interact perturbatively with the target via single gluon
exchange. And since the cross section goes as the square of the
dipole moment, one gets $\sigma_T$ proportional to $Q^2$, as is
needed.  Note however that the final state morphology is different
from what has been given for the naive parton model; it contains two
leading jets (in the virtual photon direction) and a recoil-parton
jet in the proton direction, all typically with a $p_T$ scale $Q$
(this in the collinear $\gamma^*$-proton frame; cf. Fig.
\ref{fig2}). Also the A-dependence for this mechanism is generically
$A^1$. 

The second mechanism is associated with more infrequent
configurations, where the $q$ and $\bar{q}$ created by the virtual
photon do not have large $p_T$, but are aligned along the
virtual-photon beam direction. This clearly leads to one particle
(call it the quark $q$) carrying almost all the momentum and the
other (the $\bar{Q}$) carrying much less. When the kinematics is
worked out, one finds that the typical momentum carried by the
``slow" $\bar{Q}$ is of order $x^{-1}$ GeV. But do note that this is
still hundreds of GeV for HERA conditions, in Gribov's fixed-target
reference frame.

\vspace{.5cm}

There is enough time, according to Ioffe's basic estimate, for this
``slow" $\bar{Q}$ to evolve nonperturbatively \cite{bj:springer}, and
in particular it will be found at large transverse distances from
the ``fast" quark, of order the hadronic size scale. (For example
there is enough time and space for a nonperturbative color flux-tube
to grow between $q$ and $\bar{Q}$.) So on arrival at the target
hadron the hadronic progeny of the virtual photon look something
like a B-meson, the fast pointlike $q$ analogous to the pointlike
$b$-quark, and the slow structured $\bar{Q}$ looking something like
the light constituent antiquark orbiting the $b$.

It follows that when this configuration evolves, it should be
absorbed geometrically on the target nucleus or nucleon, as assumed
by Gribov. But the probability, per incident virtual photon, that
this configuration actually occurs is easily worked out to be $({\rm
constant})/Q^2$, so the scaling of the structure function $F_2$ is
recovered. Also, the final-state structure in collinear reference
frames contains no jets, so the parton-model final state structure
is recovered. The $\bar{Q}$ final-state fragmentation products are
in fact located in the ``hole" fragmentation region already
described. An additional expectation is that the A-dependence at
small $x$ is $A^{2/3}$, even at large $Q^2$.

\section{Phenomenology}   
 
I am not conversant with all the details of the phenomenology.
However to the best of my knowledge the main features of the data
support the ``aligned jet" mechanism for the bulk of the events which
build $F_2$ at small $x$. In particular,

\begin{enumerate}
\item
The A-dependence of $F_2$ at small $x$ and large $Q^2$ is 
roughly $A^{2/3}$; 
shadowing is definitely seen and scales \cite{e665}. 

\item
Leading dijets (in the virtual photon direction) are seen rarely, if
at all, when the data is viewed in a collinear $\gamma^*$-proton
reference frame. The final-state is ``soft" most of the time
\cite{h1}.

\item
However, in the region of sharp rise of $F_2$, I am not sure that
these final-state properties persist as strongly. Certainly the
A-dependence is untested there. 
\end{enumerate}

\noindent
But in any case, it would appear that the most natural hypothesis is
that the data on $F_2$ is dominated by the aligned-jet mechanism at
small $x$.

\section{Diffraction}

With this lengthy prologue, we are now ready to consider the
rapidity-gap events seen at HERA. But with these preliminaries the
interpretation is very simple and direct. In particular, whenever
there is a process where strong absorption occurs (transmission
probability small compared to unity), there must be (elastic) shadow
scattering. In this case it is the ``slow" $\bar{Q}$ which is the
structured, hadronic object which gets absorbed. So we can expect it
to be elastically scattered from the proton (or nucleus) as well.

What will the final state look like?  The $\bar{Q}$ does not emerge
unscathed, but will physically separate from the ``fast" quark $q$.
So there will be hadronization associated with this color
separation, just as in $e^+ - e^-$ annihilation. In the lego plot of
the final state, this means that a population of hadrons will be
found between the fragmentation region of the quark $q$ and the
``hole" fragmentation region characteristic of the rapidity of the
$\bar{Q}$ before-and after-the elastic scattering; the mass of this
hadron system is typically of order $Q^2$. Hadrons will {\it not} be
found, however, in the rapidity region between the target proton (or
nucleus) and the $\bar{Q}$.

Actually the distribution in the diffracted mass $m$ can be inferred
from the Gribov estimate, Eq. 3, because the momentum change of the
$\bar{Q}$ due to the elastic scattering is typically so small that
the mass of the $q$-$\bar{Q}$ system is not significantly
modified. The Gribov distribution associated with $(1 - Z_3)$ is
\begin{equation}
     \frac{dN}{dm^2} = \frac{m^2}{(Q^2~+~m^2)^2}\ .
\end{equation}

However this should be multiplied by the alignment probability
${\rm (constant)}/m^2$. Experimentalists prefer to use instead of the
diffracted mass the scaled quantity beta:
\begin{equation}
      \beta = \frac{Q^2}{m^2~+~Q^2} \ .
\end{equation}
Therefore
\begin{equation}
      \frac{dN}{d\beta} \sim ({\rm const}) \ .
\end{equation}

A constant beta distribution, as estimated here, is in rough
agreement with the data \cite{hi:beta,zeus:beta}, especially given
the semiquantitative nature of these arguments. However there does
appear to be an excess at small beta (large diffracted mass), which
requires an extension of this mechanism such as inelastic
diffraction of the constituent quark.

The other dependence of relevance is that of the $W^2$ dependence of
the ratio of the diffractive component to the total. It should be
(at fixed $Q^2$) the same as the $s$-dependence of
$\sigma_{el}/\sigma_{tot}$ for hadron-hadron interactions. Donnachie
and Landshoff \cite{dl} successfully fit the total and elastic cross
section data with a Pomeron Regge pole, namely a pure power-law
dependence of $\sigma_{tot}$. The behavior is $s^{0.08}$. This
should also be the case (up to a logarithm associated with the
shrinkage of the elastic peak) for $\sigma_{el}/\sigma_{tot}$. The
$W^2$ dependence of $F_2$ in this picture (at fixed $Q^2$) should
also be $(W^2)^{0.08}$; this number seems on the low side 
\cite{hi:beta,zeus:beta} but there
is still controversy and uncertainty on what the fixed-$Q^2$
exponent really is. In any case, the fraction of rapidity-gap events
should not be a strong function of either $Q^2$ or $W^2$.

Finally, the absolute magnitude of the ratio of gap/no-gap events,
predicted to be $\sigma_{el}(\bar{Q}-p)/\sigma_{tot}(\bar{Q}-p)$, is
reasonable \cite{h1,ahmed}---between 5\%\ and 10\%.

Omitted in this line of argument, but certainly possible to include,
are the contributions of diffraction-dissociation of $\bar{Q}$
and/or target proton/nucleus. Although a year ago \cite{aaaa} I
essentially assumed (in the language of this talk) that the
$\bar{Q}$ diffraction dissociation might be dominant, this year it
seems more unreasonable---especially in the light of the data that
appeared in the intervening time. It does seem that excitations of
constituent quarks are not seen in spectroscopy, and that may be
reflected in the HERA diffractive data. The fraction of rapidity-gap
data for which the proton dissociates should be more or less
characteristic of the ratio of single dissociation to elastic
scattering (25\%\ or so) seen in $p-\bar p$ collisions. This should
be soon checked at HERA.

It certainly is possible to sharpen this line of thinking and make
more crisp predictions \cite{fs}.
And the recent ideas of Buchmuller and
Hebecker \cite{bh} bear much similarity to this picture. I apologize
for not having done more myself. But the bottom line of this line of
thinking, worth emphasizing here again, is that the important
mechanism for the small-$x$ final-state structure is not to be found
within perturbative QCD. It is not short-distance, weak-coupling
dynamics that counts, but large-distance, strong-coupling,
strong-absorption dynamics that is at the heart of the matter. There
need be nothing more pointlike about the mechanism producing the
diffractive final states than the mechanism responsible for elastic
proton-proton scattering.

\section{What about (BFKL) Hard Diffraction?}

The first mechanism for the small-$x$ dynamics which was discussed
in Section 2, i.e. ``color-transparency" or ``QCD Bethe-Heitler" (or
noncollinear photon-gluon fusion), must at some level also be
present. For the reasons already cited, I suspect it is at no more
than the 10\%--20\%\ level. But that is only a guess. The best way
to isolate it experimentally is via the 3-jet final state morphology
exhibited in Fig. \ref{fig2}. This is the seed kernel for building
at high energies and fixed $Q^2$ the BFKL $W^2$ dependence
\cite{bfkl} via production of extra gluons into the phase space,
gluons typically also carrying $p_T$ of order $\sqrt{Q^2}$. The
$W^2$ dependence to be expected is much stronger, of order
$(W^2)^{0.4}$.

Open questions regarding the relevance of this mechanism for HERA
include 
\begin{enumerate}

\item
whether the normalization of the lowest-order kernel is large enough,

\item
how much room there is in the available HERA phase space for
building up the power-law behavior, and

\item
whether the scheme is consistent: there exist criticisms regarding
``diffusion into the infrared", as well as claims
\cite{cd,vdel,boa,bj} that more careful attention must be paid to
energy-conservation constraints within the multi-Regge kinematics.
\end{enumerate}

\vspace{.5cm}
\begin{figure}[htbp]
\begin{center}
\leavevmode
\epsfxsize=5in
\epsfbox{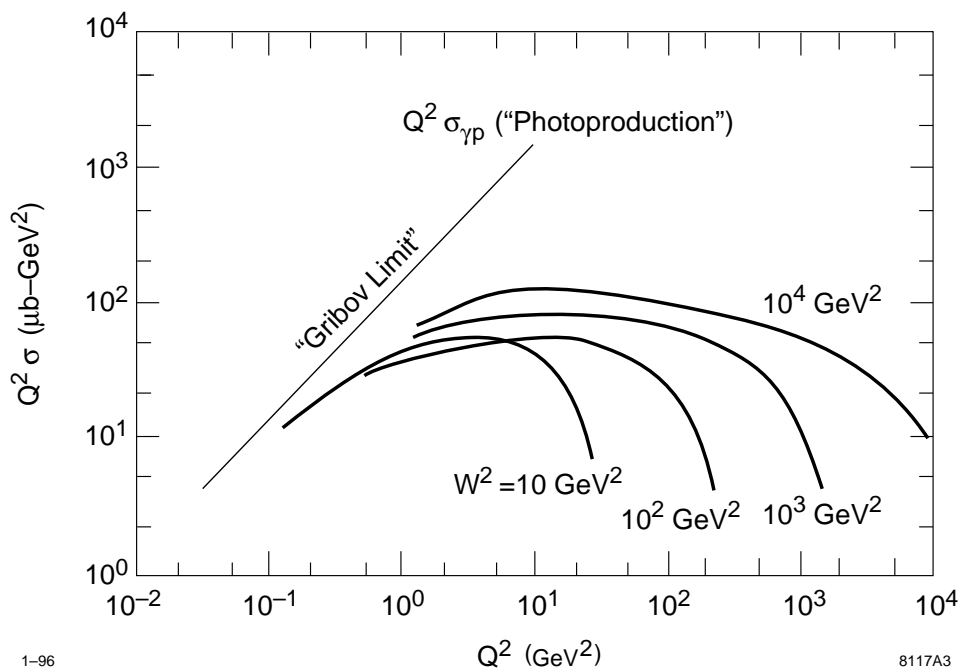}
\end{center}
\caption[*]{A log-log sketch of $F_2$ vs $Q^2$ at various fixed values
of $W^2$.}
\label{fig3}
\end{figure}

\noindent
An observed trend toward the BFKL $W^2$ dependence would clearly be
of fundamental importance, implying a new class of nonperturbative,
absorptive effects going far beyond those we used in the previous
section to interpret existing data. I here make only the most modest
suggestion, regarding how to plot the data. I am a firm believer in
the importance of searching for the optimum way of presenting data,
the way which most directly highlights what is important. My
suggestion in this case is to plot log $F_2$ versus log $Q^2$ at
fixed $W^2$. A sketch of what I mean is shown in Fig. \ref{fig3}.
$W^2$ is chosen rather than $x^{-1}$ because there is no longer
scaling in the small-$x$ region, and because the nonscaling depends
on gluon emission, which probably is more dependent on the amount of
available phase space than anything else. (This is certainly the case
for BFKL). The logarithmic scales allow a clear view of how the
photoproduction limit is approached, and above that limit by some
not-so-well-defined factor is the Gribov bound, unmodified by any
damping due to color-transparency or aligned-jet configurations.
Existing data for not large $W^2$ show a curve of log $F_2$ vs log
$Q^2$ which is concave down. If any part of the curve becomes at
high $W^2$ concave up, this would be to me a signal for ``BFKL
behavior", because it is the only way the Gribov bound can be
reached. The important regime for HERA is, as is well-known,
moderate $Q^2$ (0.5 GeV$^2$--15 GeV$^2$) at the highest $W^2$
attainable.

My own favorite guess \cite{bj:vietri} on how things will turn out
is that the curves will remain concave down, but that the $W^2$
dependence of the maximum of $F_2(Q^2, W^2)$ for given $W^2$ will
behave more or less like BFKL.

\section{Acknowledgment}

It has been some time since I last visited ITEP, and as always it
has been a most pleasant and stimulating experience. I thank Boris
Ioffe and his colleagues for their warm hospitality.


\newpage

\begin{thebibliography} {99}

\bibitem{wagner} 
F. Wagner, these proceedings.

\bibitem{mueller} 
A. Mueller, these proceedings.

\bibitem{lipatov} 
L. Lipatov, these proceedings.

\bibitem{bj:align} 
J. Bjorken, AIP Conference Proceedings No. 6, Particles and Fields
Subseries No. 2, ed. M. Bander, G. Shaw, and D. Wong (AIP, New
York, 1972).

\bibitem{ioffe} 
B. Ioffe, Phys. Lett. {\bf 30}, 123 (1968).

\bibitem{gribov} 
V. Gribov, Sov. Phys. JETP {\bf 30}, 709 (1969).

\bibitem{bj:holes} 
J. Bjorken, Phys. Rev. {\bf D2}, 282 (1973). 

\bibitem{bj:springer} 
J. Bjorken, Lecture Notes in Physics 56, ``Current Induced
Reactions," ed. J.~K\"orner, G. Kramer, and D. Schildknecht,
Springer-Verlag (Berlin, 1976), p. 93.

\bibitem{e665}
M. Adams, {\it et al.,} (E665 Collaboration), Z. Phys. {\bf C67},
403 (1995) and references therein.

\bibitem{h1}
M. Derrick {\it et al.,} (Zeus Collaboration), Phys. Lett. {\bf
B332}, 228 (1994).

\bibitem{hi:beta}
T. Ahmed {\it et al.,} (H1 Collaboration), Phys. Lett. {\bf B348},
681 (1995).

\bibitem{zeus:beta}
M. Derrick {\it et al.,} (Zeus Collaboration), Z. Phys. {\bf C68},
569 (1995).

\bibitem{dl} 
A. Donnachie and P. Landshoff, Nucl. Phys. {\bf B267}, 640 (1986).

\bibitem{ahmed}
T. Ahmed {\it et al.,} (H1 Collaboration), Nucl. Phys {\bf B429},
477 (1994).

\bibitem{aaaa}
J. Bjorken, {\it International Workshop on Deep Inelastic
Scattering and Related Subjects,} Eilat, Israel, 6--11 February
1994, ed. A.~Levy (World Scientific, Singapore, 1994), p. 151.
  
\bibitem{fs} 
See the excellent discussion of much of this material by H.
Abramowicz, L. Frankfurt, and M. Strikman, DESY preprint DESY-95-047
(March 1995), where further references can be found.

\bibitem{bh} 
W. Buchmuller and A. Hebecker, Phys. Lett. {\bf B355}, 573 (1995).

\bibitem{bfkl} 
E. Kuraev, L. Lipatov, and V. Fadin, Sov. Phys. JETP {\bf 44}, 443
(1976); Y.~Balitsky and L. Lipatov, Sov. J. Nucl. Phys. {\bf 28},
822 (1978). 

\bibitem{cd} 
J. Collins and P. Landshoff, Phys. Lett. {\bf B276}, 196 (1992). 

\bibitem{vdel}
V. Del Duca and C. Schmidt, Phys. Rev. {\bf D49}, 4510 (1994).

\bibitem{boa} 
B. Andersson, {\it Proceedings of the XVIV International Symposium
on Multiparticle Dynamics,} Vietri sul Mare, Salerno, Italy, 12--19
September 1994, ed. A.~Giovannini, S. Lupia, and R. Ugoccioni (World
Scientific, Singapore, 1995), p. 263.

\bibitem{bj}
J. Bjorken, {\it Proceedings of the XVIV International Symposium on
Multiparticle Dynamics,} Vietri sul Mare, Salerno, Italy, 12--19
September 1994, ed. A. Giovannini, S. Lupia, and R. Ugoccioni (World
Scientific, Singapore, 1995), p. 579.

\bibitem{bj:vietri} 
J. Bjorken, SLAC preprint SLAC-PUB-95-6949, to be published in the
{\it Proceedings of the Lake Louise Winter Institute on Quarks and
Colliders}, Lake Louise, Canada, February 19--25, 1995. 

\end{thebibliography}
\end{document}